**Title: Electrically controllable router of interlayer excitons**

**Authors**

Yuanda Liu,[1, 2] Kostya S. Novoselov[3*] and Weibo Gao[1, 2*]

**Affiliations**

[1] Division of Physics and Applied Physics, School of Physical and Mathematical Sciences, Nanyang Technological University, Singapore 637371, Singapore.

[2] The Photonics Institute and Centre for Disruptive Photonic Technologies, Nanyang Technological University, Singapore.

[3] Department of Material Science & Engineering, National University of Singapore, 117575, Singapore.

Corresponding author. Email: kostya@nus.edu.sg and wbgao@ntu.edu.sg

**Abstract**

Optoelectronic devices which allow rerouting, modulation and detection of the optical signals would be extremely beneficial for telecommunication technology. One of the most promising platforms for such devices are excitonic devices, as they offer very efficient coupling to light. Of especial importance are those based on indirect excitons, because of their long lifetime. Here we demonstrate excitonic transistor and router based on bilayer of WSe2. Due to their strong dipole moment, excitons in bilayer WSe2 can be controlled by transverse electric field. At the same time, unlike indirect excitons in artificially stacked heterostructures based on transition metal dichalcogenides - naturally stacked bilayer offers long exciton lifetime, smaller non-radiative losses, and are much simpler in fabrication.

**MAIN TEXT**

**Introduction**

Increasing demand for faster telecommunication technologies calls for the shift of signal processing from electronic to optical domain. A very promising opportunity in this area is provided by excitonic devices.(*1-5*) Such devices convert light into excitons, manipulate excitons by means of electric or magnetic fields and convert excitons back to light. Of particular importance are devices based on indirect excitons, which offer much longer (up to two orders of magnitude) lifetime in comparison with the direct excitons.(*6-8*)

Originally designed on the type-II quantum wells in GaAs/AlGaAs heterostructures, such devices indeed offered extended lifetime of indirect excitons.(*3-5, 9*) With the advent of 2D materials and especially transition metal dichalcogenides (TMDC) it became possible to form new types of type-II quantum wells by combining TMDCs of different chemical compositions.(*10-18*) Such materials are potentially more promising for optoelectronic devices due to larger exciton binding energy (hundreds of meV), which should enable them to be operational at elevated temperatures. Indeed, it has been demonstrated that the optoelectronic devices based on stacked monolayers of TMDC can be operational at room temperatures. (*1, 19*) However, stacking of TMDCs to form van der Waals heterostructures is a tedious process, which often results in contamination of the interface and leads to non-radiative losses. Furthermore, the crystallographic directions of the stacked crystals have to be aligned to minimize momentum mismatch between electron and hole in such indirect exciton, which presents a significant technological problem should such devices go to applications.(*20-25*)

Here we propose to use naturally occurring 2H stacked bilayer of WSe2 in order to produce indirect excitons. Electrons and holes get separated to the different layers, forming spatially indirect excitons. Unlike mechanically stacked van der Waals heterostructures based on different layers of TMDC, our naturally stacked bilayer WSe2 doesn't have any contaminants between the layers, thus providing a very high quality of heterostructures and very long lifetime of the indirect excitons. The potential energy and binding energy can be controlled by the transverse electric field, which allows us to manipulate such indirect excitons remotely and demonstrate transistor and router behaviour.

**Results**

Our samples have been produced from 2H bounded bilayer WSe2 by stacking it between two thick (approximately 20 nm) hexagonal boron nitride (hBN) crystals (see Methods for the details of the sample preparation). The whole stack was transferred on Si/SiO2 substrate with prepatterned gold electrodes (the particular shape of the electrodes depends on the particular experiment). A global graphene gate has been transferred on top of the whole stack. All measurements presented in this work were performed with two devices. Device 1 has bottom electrodes shaped as a series of parallel stripes (Fig. 1, as well as Fig. S1, S2, S5 and S6). The geometry of device 2 is designed to be three-beam star, Fig 3.

We first measured PL (photoluminescence) spectra and spatial emission to characterize device 1. Bilayer WSe2 is an indirect band gap semiconductor due to the finite interlayer coupling. Its valence band maximum and conduction band minimum are located at the K(K') and Q(Q') points of the Brillouin zone, respectively. Therefore, the lowest-energy optical transitions in bilayer WSe2 are indirect transitions between the K and Q valleys (*6, 26*). We note here that the upper layer and bottom layer of bilayer WSe2 has a different surrounding environment (with hBN and graphene, or hBN and gold electrode). This will shift the binding energy of excitons in upper layer and bottom layer (*27*), and therefore shift the conduction and valence band to form the type-II band alignment. Holes will be in the bottom layer and electrons will be in the upper layer to form the interlayer excitons. Figure 1A shows that the interlayer exciton emission has a peak wavelength at ~784 nm, which agrees with K-Q (k'-Q') transitions. The fixed orientation of exciton dipole direction also agrees with experiment demonstrated below. The wavelength does not overlap with the intralayer exciton emission (~736 nm)(*6, 28*). This permits us to remove the intralayer exciton emission using optical longpass filters.

We characterize the exciton wavelength and lifetime as shown in Figure 1B and Fig. S2. Electric field (Figure 1B) shows continuous wide tunability of exciton wavelength, which is actually a combined effect of binding energy tuning, Coulomb scattering and screening of the Coulomb interaction, as well as change of repulsive interaction with electric field (*29*). Time-resolved PL measurements (Fig. S2) reveal a lifetime well in the nanosecond time range, representing two order-of-magnitude improvement over that of intralayer excitons, allowing formation of a long-lived 2D exciton gas.(*6*) Within this lifetime, IXs travel over a long distance larger than the device dimensions. Electric-exciton interaction controls 2D potential for IXs in atomically thin homogeneous material, as schematically illustrated in Figure 1C.

We demonstrated 2D potential-energy controllability of interlayer excitons by an out-of-plane electric field. An interlayer exciton has a dipole moment p with an out-of-plane (z) direction due to the separation and confinement of composed electron and hole in spatially separated layers. Therefore, application of an out-of-plane electric field Fz leads to exciton energy shift(*2, 30, 31*) of ΔE = *ed*Fz, where *ed* is the built-in dipole moment of IXs, and d is close to the distance between the monolayer centers (~0.78 nm for WSe2 bilayer, as measured by AFM. Fig. S3). With a fixed exciton dipole direction, the electric field can therefore create lower or higher energy potential along the electrode and therefore drives the confinement and spread of excitons.

The excitons will be driven towards regions of lower energy in the 2D potential energy surface. We apply voltage to one of the bottom electrodes, while the top graphene is grounded. Therefore, a negative voltage creates a downwards vertical electric field that lowers the exciton electrostatic potential by ΔE = edFz (top panel in Figure 1D), yielding a potential well overlapping the area of the bottom electrode. The experimental measurement of the exciton spatial movement is performed using the set-up shown in Fig. S4. The real-space emission map (Figure 1E) shows unambiguously exciton confinement and diffusion along the electrode with an elongated shape.

On the contrary, when the exciton potential is increased by a positive voltage (bottom panel in Figure 1D), the excitons are spreading away from the anti-channel. The real-space emission map for the spreading is displayed in Figure 1G. The emission trace extends in a direction perpendicular to the electrode rather than a circular cylinder around the laser spot (free diffusion shown in Figure 1F), indicative of realization of a long-range excitonic unidirectional movement driven electrically. Note that the position of the emission maximum shifts away from the illumination point to the side of the electrode, benefitting from the long lifetime of IXs. Figure 1H plots the emission intensity profiles along the flux trace extracted from Figure 1E and 1G, showing clearly the transition from trap regime to the spreading regime by modulating the gate voltage. An emission intensity valley at the excitation spot implies the highly tunable potential. Fig. S5 exhibits the real-space emission map of excitons for gate voltage $V_G$ = - 14 V ~ 14 V at a step of 1 V. For device 2, when excitation is on the gate electrode and gate voltage applied to that electrode, similar confinement and spreading is observed and are shown in Fig. S6 for PL emission pattern difference between $V_G$ = 10V and -10V.

The exciton flux can be controlled by an electric field in the prorogation midway, as exhibited in Figure 2. The laser light was illustrated on the Source (S)-electrode area. The generated excitons propagate away from the excitation site. When a negative voltage $V_G$ was applied to the G electrode, an exciton potential well is created in the gate electrode area, resulting in the termination of the exciton flux. Only noise signal can be detected on the Drain (D) electrode, as shown in Figure 2A. This agrees with the exciton transistor as demonstrated with GaAs double quantum wells (*4*). A positive voltage of $V_G$ applied to the G electrode can separate emission area on the source electrode and drain electrode, but will not fully block the diffusion of exciton flux (Figure 2B). We therefore can define $V_G$ = 10V as ON state and $V_G$ = -10V as OFF state for the intensity on the drain electrode. Fig. S7 shows the real-space emission map of excitons for G voltage $V_G$ = - 9 V ~ 9 V at a step of 2 V.

**Excitonic transistor operation** Next, we demonstrate excitonic switching on the basis of unidirectional movement of excitons. An excitonic circuit composed of three exciton transistors was fabricated (See methods for detailed fabrication process). Figure 3A shows the optical image of the circuit. Excitonic transistor is designed to be a device with three terminals including source, gate, and drain, as schematically shown in Figure 1C. It operates like an electric field-effect transistor (*32-34*), which works by modulating the electrical charge carrier density, and hence electrical resistance of a thin semiconducting channel through the application of an electric field using gate. The exciton intensity difference drives the IXs generated under excitation to move from the source towards drain. Application of a gate voltage electrostatically modulates the prorogation distance and concentration. A negative voltage traps the IXs in the region of the gate, thereby lessening the IXs concentration propagating to drain. In our design, three transistors have a common electrode, the triangle electrode as shown in Figure 3A.

We examine the net effects of gate voltage on the IXs population distribution. Figure 3B, 3C present real-space emission maps when illustrating at the center of triangle electrode. To this end, the net increment of the emission intensity is the difference of PL emission when transistor is ON state (gate electrode 10V) and OFF state (-10V). This also avoids the noise of the background induced by the measuring system. The emission image for one transistor in ON state is shown in Figure 3A. The $V_S = 0V$ and $V_D = -1$ V create a potential energy offset, and the perturbation in gate voltage $\Delta V_G = 20$ V. In the device, the distance between the source and drain is 4 um. At the same time, the other two transistors are kept inactive. The star geometry of the circuit allows star switching. Figure 3C shows that transistor 2 is also switched on. The IXs flux is routed simultaneously on two paths with an angle of 120º.

**Excitonic router operation** We further reconfigure the same circuit to demonstrate point-to-point movement to implement device functionality of exciton circulation. We consider the three outer electrodes as the three ports of a router, which are denoted by port 1, 2, and 3 in Figure 4, respectively. Firstly, excitation laser spot is focused on port 1. Potential energy difference is created by the application of voltages on drain electrode port 1 (0 V) and port 2 (-3 V), while applying a high voltage (3 V) on port 3. Such a regime allows excitons generated at port 1 to only be transmitted to port 2. Figure 4B and 4C show the exciton propagation from port 2 to only port 3 and, likewise, from port 3 to only port 1. Figure 4D displays the excitons move from port 3 to port 1 and port 2 simultaneously. The excitons transmit either to one port or two port depending on the voltage application. Therefore, the router is programmable, and its operation direction is defined by the electric voltages applied on each port. The essential ingredients of our excitonic router are the all-optical input-output signal processing, the miniaturization of device dimensions, and the flexibility of low-voltage electrical driving scheme.

**Discussion**

In summary, our experiments demonstrate electrical controllability of exciton dynamics in atomically thin homogeneous materials, presenting exciton propagation in a controlled direction. This enables conceptual design of miniaturized on-chip all-optical devices, the excitonic switching and the excitonic routing in an integrated circuit. We envision that this proof-of-concept device may pave a promising way for developing practical exciton-related applications, and inject new vitality into research in this bosonic particle.

**Materials and Methods**

**Sample Fabrication**

The exciton devices are composed of the bottom electrodes, two hBN (hexagonal BN) gate dielectric layers, a WSe2 bilayer, and a graphene homogeneous top gate. The bottom electrodes were patterned using standard electron beam lithography followed by sputtering of Au (~10 nm) onto 285 nm thick SiO2/Si substrate. Intrinsic silicon was adopted to considerably decrease the background emission intensity. WSe2 bilayer, hBN, and graphene were mechanically exfoliated from bulk single crystals and were subsequently stacked by means of dry transfer onto the bottom electrodes.(*35*) The layer number of WSe2 was identified by a combination of optical contrast, PL spectroscopy and AFM (atomic force microscopy). A second step of electron beam lithography was performed to define the top gate metallic contacts (Ti/Au = 3 nm/60 nm) followed by electron-beam evaporation and lift-off.

**Optical Characterization**

Device was mounted on the cold head of a closed-cycle helium cryostat with variable temperature (5 K < T < 350 K). Optical measurement was performed with a customized optical setup (see Supplementary Figure 5 for more details). We excited the sample with the above-bandgap continuous-wave lasers focused to a ~1 um spot on the surface of the sample using a 50 × long working distance microscope objective lens (numerical aperture = 0.65). The scattered photoluminescence signal was collected vertically by the same objective. Laser spot was positioned using either piezo-electric nanopositioner or a galvo mirror pair. To measure the spatial emission maps, the PL signal was de-scanned by another galvo mirror pair, and detected by a single photon detector. Filters were used in all experiments to prevent pump light from reaching any detectors or cameras. The pumping power was adjusted using a neutral density filter.

The electric voltage was applied using a Keithley 4200-SCS semiconductor parameter analyzer and a Keithley 2450 source-meter.

**H2: Supplementary Materials**

Fig. S1. Schematic structure and optical image of device 1.
Fig. S2. Time-resolved photoluminescence of WSe2 bilayer.
Fig. S3. Interlayer distance of WSe2 bilayer.
Fig. S4. Schematic diagram of the optical setup.
Fig. S5. Real-space emission map of excitons for gate voltage $V_G$ = - 14 V ~ 14 V at a step of 1 V.
Fig. S6. Real-space emission map of excitons. The intensity is the subtraction of that for $V$ = -10 V from that for $V$ = 10 V.
Fig. S7. Real-space emission map of excitons for gate voltage $V_G$ = 9 V ~ -9 V at a step of 2 V.

## Acknowledgments

**Funding:** This work is support by the Singapore NRF fellowship grant (No. NRF-NRFF2015-03) and Singapore Ministry of Education [MOE2016-T2-2-077, MOE2016-T2-1-163, and MOE2016-T3-1-006 (S)]

**Author contributions:** All authors contribute extensively to the manuscript.

**Competing interests:** The authors declare no competing interests.

**Data and materials availability:** The data that support the plots within this paper and other findings of this study are available upon reasonable request.

**Figures and Tables**

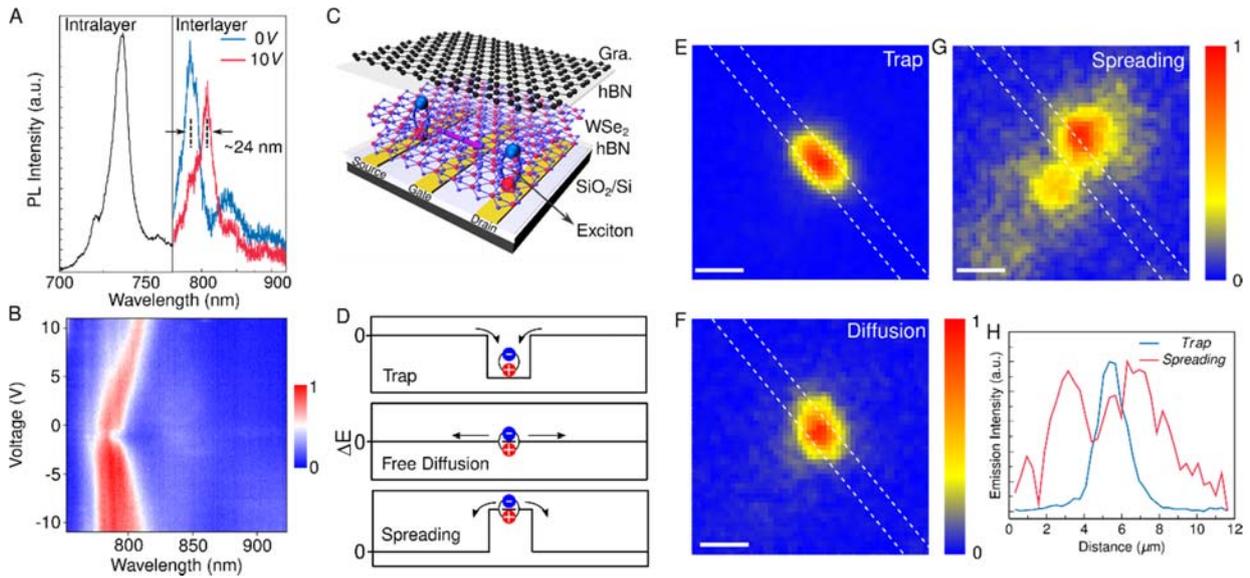

**Fig. 1. Control 2D potential-energy of interlayer excitons by electrical field.** (**A**) Photoluminescence of WSe2 bilayer. The emission centering at ~736 nm and ~784 nm are ascribed to the intralayer excitons and interlayer excitons transition, respectively. PL spectra of interlayer exciton for gate voltage of 0 V and 10 V, showing that the peak position redshifts by ~24 nm. Sample temperature is 10 K. (**B**) Contour plot of the interlayer exciton emission intensity as a function of the applied gate voltage and wavelength. The excitation power is 30 μW operated at 532 nm. (**C**) Schematic structural diagram of one excitonic transistor. (**D**) Schematic of exciton energy offset for exciton dynamic regimes of trap, free diffusion and spreading, when applying negative (top panel), zero (middle panel), and positive (bottom panel) gate voltages, respectively. (**E-G**) Real-space emission intensity map for exciton trap, diffusion and spreading, corresponding to gate voltage $V_G$ = -9 V, 0 V, 11 V, respectively. The white dashed lines indicate the bottom electrode edges. Scale bar, 2 μm. (**H**) Emission intensity profiles across the electrode extracted from the real-space emission intensity maps. The excitation spot was focused on the center of the electrode. The excitation power is ~ 200 nW operated at 532 nm.

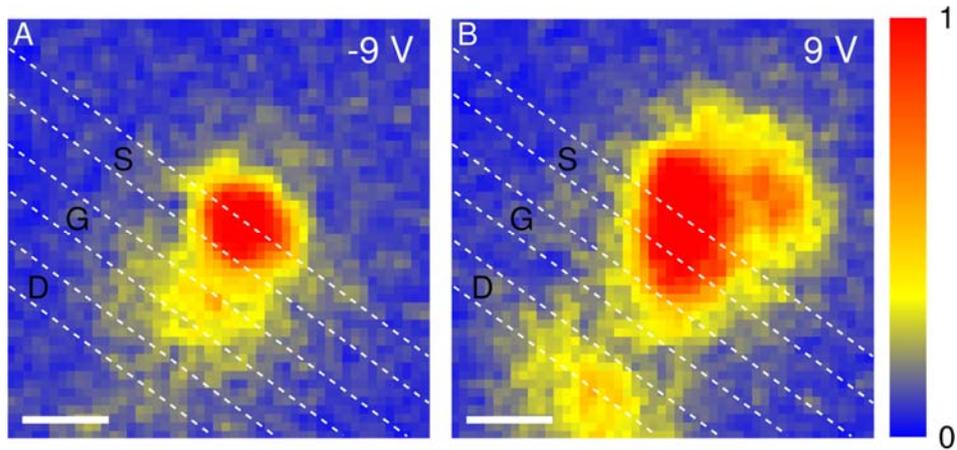

**Fig. 2. Real-space emission intensity map for exciton ON and OFF operation, corresponding to voltage applied to gate (G) electrode $V_G$ = -9 V (A), 9 V (B), respectively.** The voltage applied to Source (S) and drain (D) electrode is 4 V and 0 V, respectively. The white dashed lines indicate the bottom electrode edges. Scale bar, 2 μm. The excitation power is ~ 600 nW operated at 532 nm. Both $V_G$ = 0 V and 9V can be seen as ON state, which is in stark difference with $V_G$ = -9 V (OFF state).

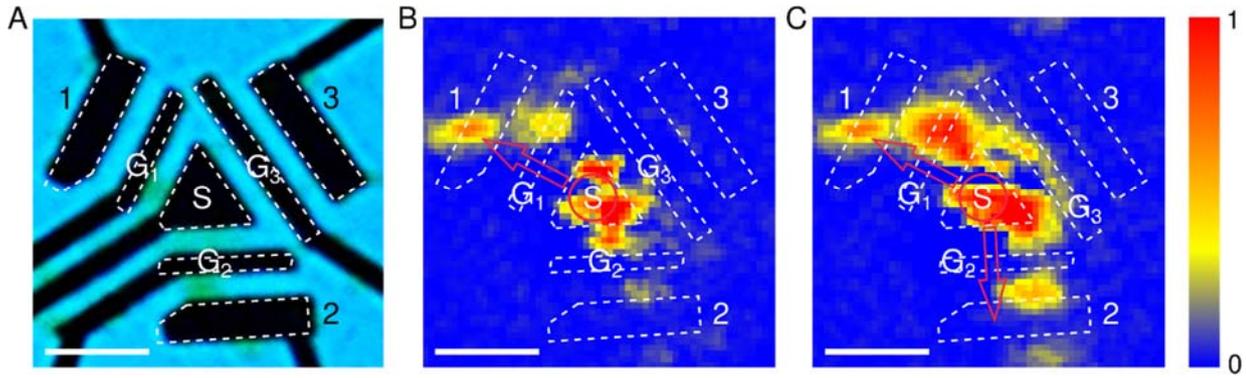

**Fig. 3. Unidirectional movement for excitonic transistor operation.** (**A**) Optical image of the device. The black areas are the bottom electrodes. (**B**) Real-space emission intensity map when one transistor is in ON state. The $V_S = 0$ V and $V_{D1} = -1$ V create a potential energy offset, and the perturbation in gate voltage $\Delta V_{G1} = 20$ V. The image shows the PL intensity difference for ON and OFF state when $V_{G1, ON} = 10$ V, $V_{G1, OFF} = -10$ V. No voltages applied on transistor 2 and 3, and they are inactive. (**C**) Real-space emission intensity map when switching on the second transistor. $V_S = 0$ V, $V_{D2} = -1$ V, $V_{G2ON} = 10$ V, $V_{G2OFF} = -10$ V. Voltages applied on transistor 1 is same as that in Figure 3(B). No voltages was applied on transistor 3. The white dashed lines indicate the bottom electrode edges. The excitation spot was focused on the center of the triangle electrode. The solid red arrows are a guide to the eye. The excitation power is ~ 100 µW operated at 730 nm. Scale bar, 5 µm.

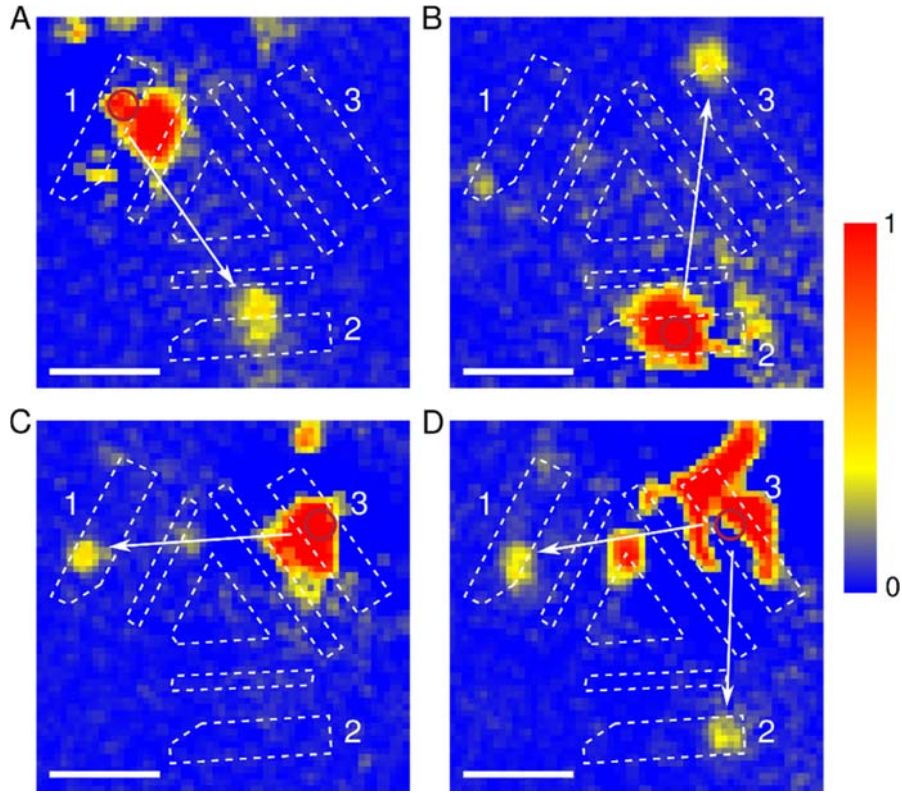

**Fig. 4. Point-to-point movement for excitonic router operation.** The exciton signal routing can be achieved by directly applying electric gate voltage on port 1, 2 and 3. All gate voltage is set to zero. (**A**) Exciton propagation from port 1 to port 2, $V_1 = 0$ V, $V_3 = 3$ V, $V_{2,\,ON} = -3$ V, $V_{2,\,OFF} = 3$ V; (**B**) Exciton propagation from port 2 to port 3, $V_1 = 8$ V, $V_2 = 0$ V, $V_{3,\,ON} = -1$ V, $V_{3,\,OFF} = 10$ V. (**C**) Exciton propagation from port 3 to port 1, $V_2 = 4$ V, $V_3 = 0$ V, $V_{1,\,ON} = -0.5$ V, $V_{1,\,OFF} = 6$ V. (**D**) Excitons propagation from port 3 to both port 1 and port 2. $V_3 = 0$ V. $V_{1,\,ON} = 0$ V, $V_{1,\,OFF} = 8$ V and $V_{2,\,ON} = 0$ V, $V_{2,\,OFF} = 8$ V. The solid white arrows parallel to the excitons propagation direction are drawn as a guide to the eye. The red circles show the laser excitation spot. The excitation power is ~ 200 μW operated at 730 nm. Scale bar, 5 μm.

**Supplementary Materials**

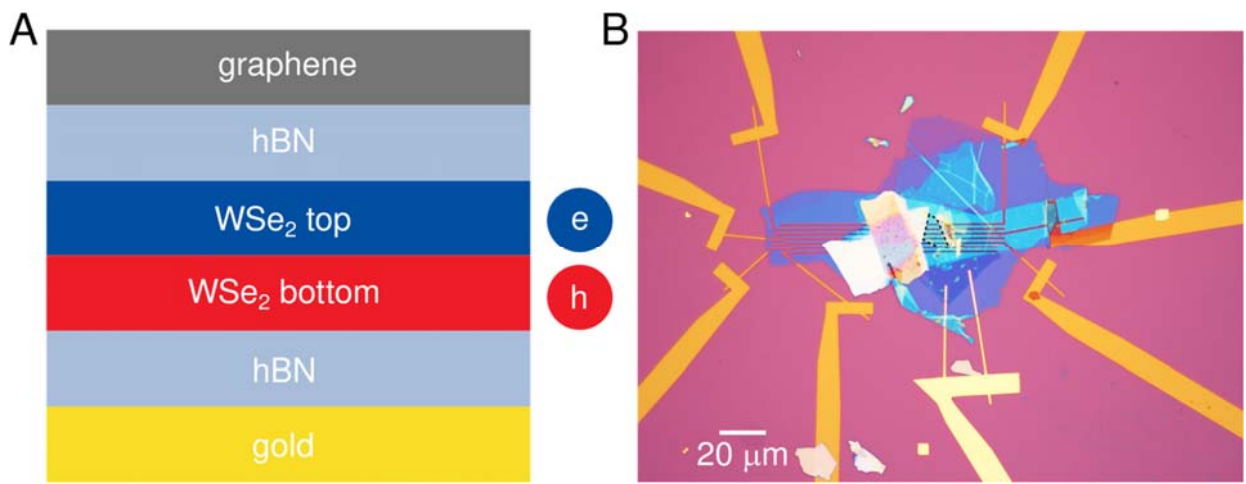

**Fig. S1. (A)** Schematic structure of device 1. **(B)** Optical image of device 1. The dashed black line shows the WSe2 bilayer area. Scale bar: 20 μm.

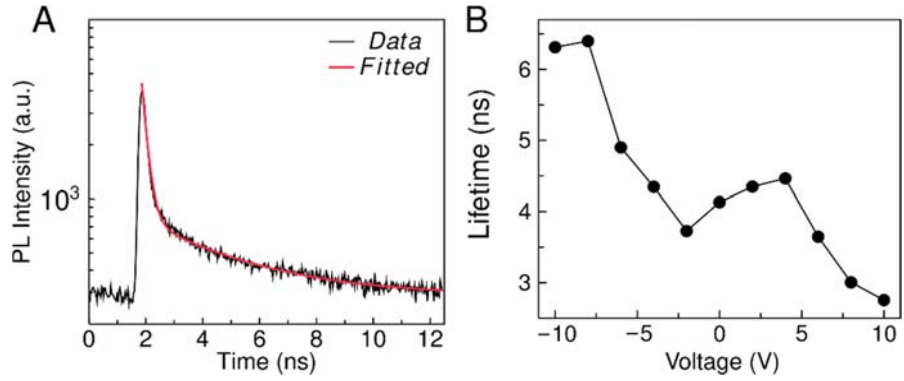

**Fig. S2. (A)** Time-resolved photoluminescence of WSe2 bilayer. $V_G = 0$ V. The black line is experimental data. The red line is fitting result by double exponential decay functions, yields two exciton lifetimes of 4.13 ns and 230 ps that can be attributed to interlayer exciton and Auger process, respectively. **(B)** Extracted lifetime values of interlayer exciton as a function of gate voltage. The sample temperature is 10 K. A diode laser operated at 532 nm with a repetition rate of 10 MHz and pulse duration of 32 ps was employed. The PL signal was detected by the time-correlated single photon counting method using the single photon detector connected to a HydroHarp system (PicoQuant). The laser average power was 2 µW.

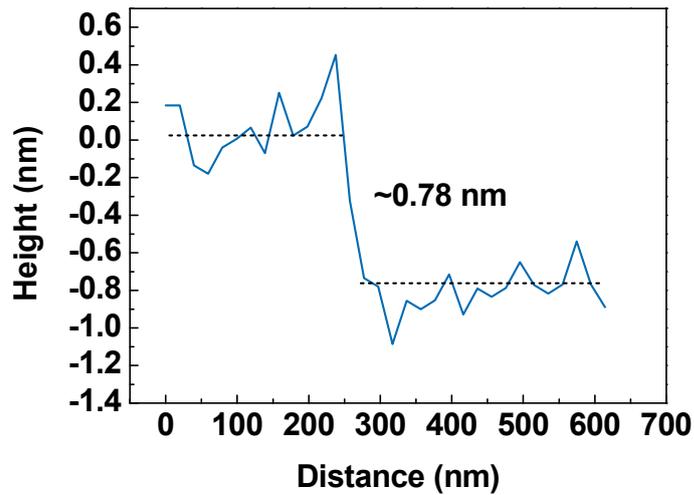

**Fig. S3.** The layer distance is measured to be ~0.78 nm, providing an evidence that the active material is WSe2 bilayer.

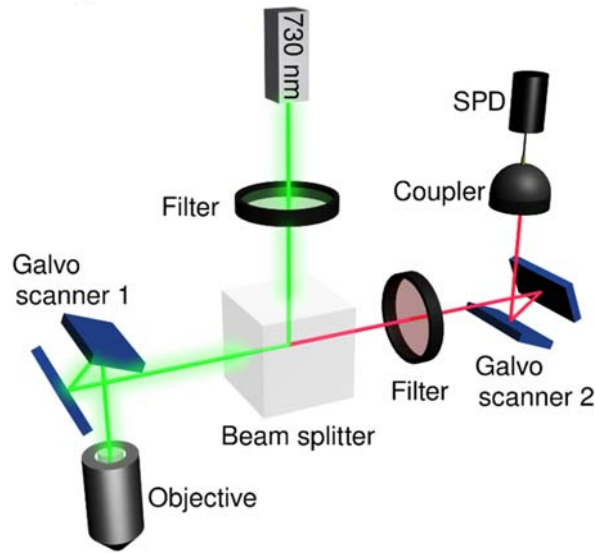

**Fig. S4.** Schematic diagram of the optical setup showing all the major optical components. The excitation laser beam was fed into scanning beam optics, which consists of rotating mirrors and focuses the laser beam onto the back of an objective lens (50 x). As the scanning mirror rotates, the beamspot moves over a wide range on the sample without aberration, allowing for high-resolution spatial scanning. To perform spatial emission scanning, galvo scanner 1 tunes the position of the laser spot, while emission signal was de-scanned by galvo scanner 2, and detected by a single photon detector.

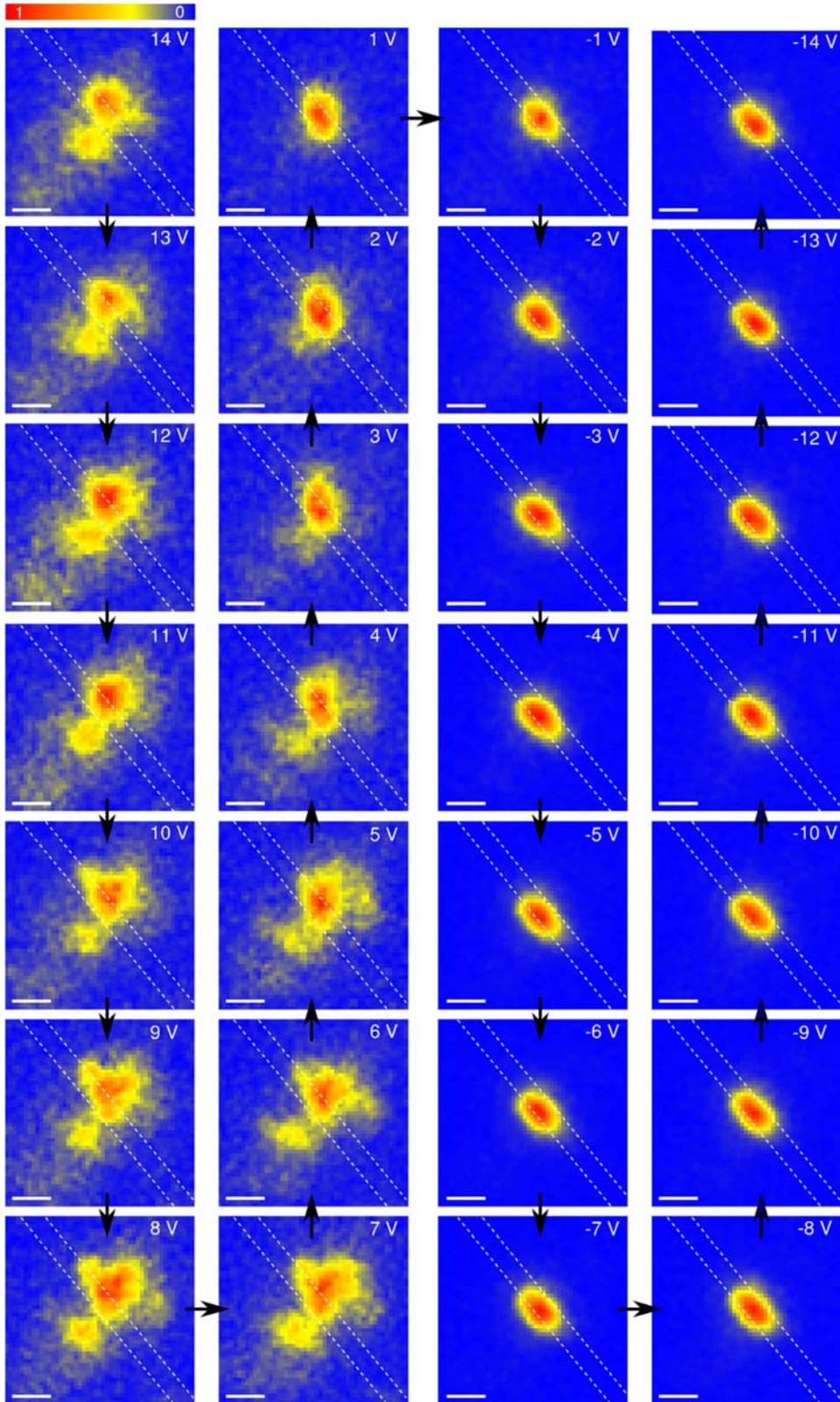

**Fig. S5**. Real-space emission map of excitons for gate voltage $V_G$ = - 14 V ~ 14 V at a step of 1 V. The white dashed lines indicate the bottom electrode edges. The excitation spot was focused on the center of the electrode. Scale bar, 2 μm.

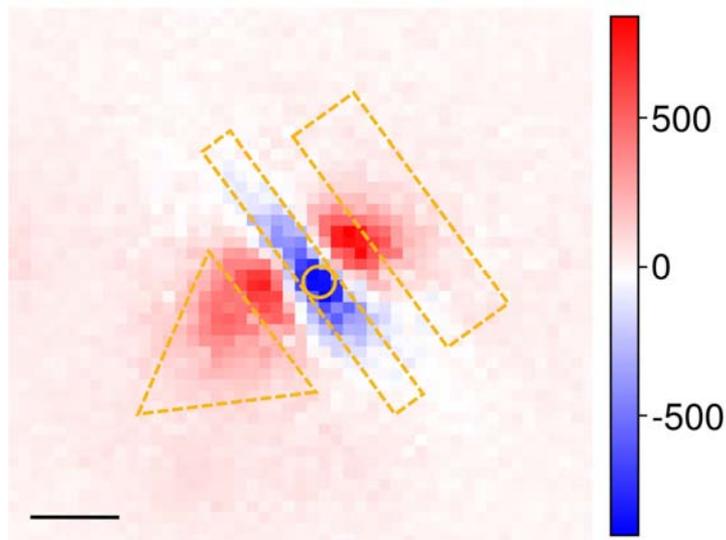

**Fig. S6. Real-space emission map of excitons.** The intensity is the subtraction of that for $V$ = -10 V from that for $V$ = 10 V, showing the confinement for $V$ = -10 V and spread for $V$ = 10 V. The voltage is applied on the middle electrode. The yellow dashed lines indicate the bottom electrode edges. The solid circle shows the position of the laser excitation. The excitation power is ~ 200 uW operated at 730 nm. Scale bar, 2 μm.

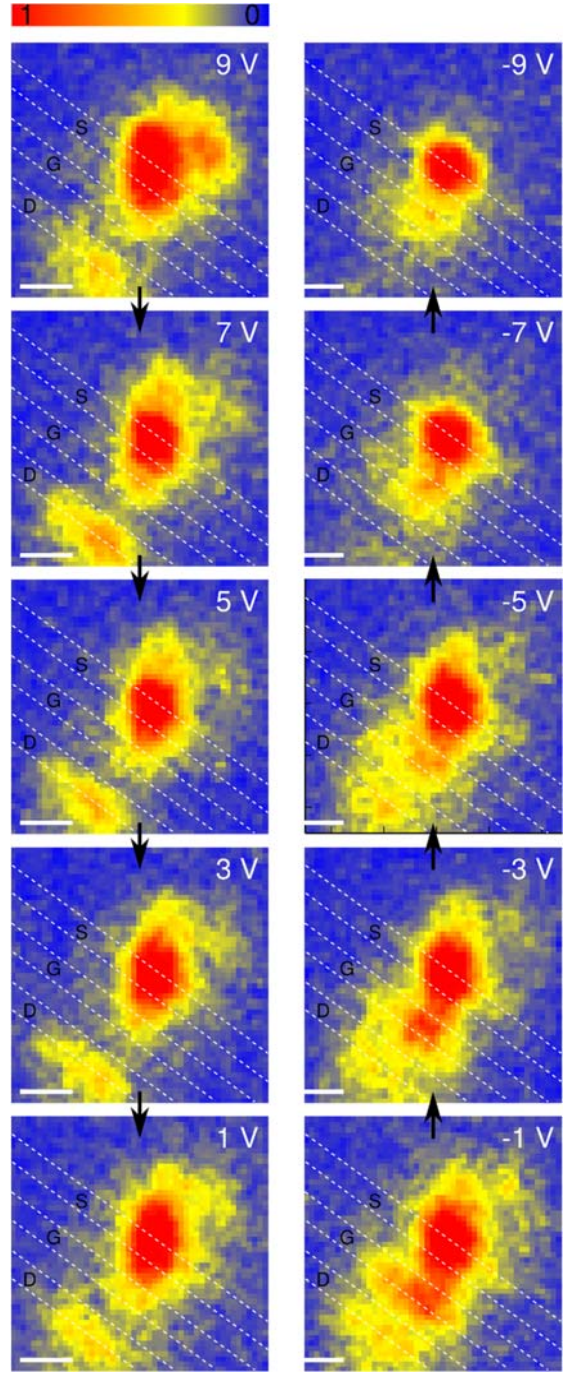

**Fig. S7.** Real-space emission map of excitons for voltage applied on G electrode $V_G$ = 9 V ~ -9 V at a step of 2 V. The voltage applied on S and D electrode is 4 V and 0 V, respectively. The white dashed lines indicate the bottom electrode edges. Scale bar, 2 μm. The excitation power is ~ 600 nW operated at 532 nm.